\documentstyle[twocolumn,prl,aps,psfig]{revtex}
\begin{document}
%\baselineskip .6cm
%\draft
\def\be{\begin{equation}}
\def\ee{\end{equation}}
\def\bfi{\begin{figure}}
\def\efi{\end{figure}}
\def\bea{\begin{eqnarray}}
\def\eea{\end{eqnarray}}
\title{Fast growth at low temperature in vacancy-mediated phase-separation}
\author{Claudio Castellano}
\address{
Fachbereich Physik, Universit\"at GH Essen, 45117 Essen, Germany and\\
Istituto Nazionale di Fisica della Materia,
Unit\`a di Roma 1, Universit\`a "La Sapienza",
piazzale A. Moro 2, 00185 Roma, Italy}
\author{Federico Corberi}
\address{Istituto Nazionale di Fisica della Materia,
Unit\`a di Salerno and Dipartimento di Fisica, Universit\`a di Salerno,
84081 Baronissi (Salerno), Italy}

\maketitle
 
\begin{abstract}

We study the phase-separation dynamics of a two-dimensional 
Ising model where A and B particles can only exchange position with a vacancy.
In a wide range of temperatures the kinetics is dominated,
during a long preasymptotic regime,
by diffusion processes of particles along domain interfaces.
The dynamical exponent $z$ associated to this mechanism 
differs from the one usually expected for Kawasaki dynamics and
is shown to assume different values depending on temperature and 
relative AB concentration.
At low temperatures, in particular, domains grow as 
$t^{1/2}$, for equal AB volume fractions.

\end{abstract}
 

When binary alloys are quenched into the unstable region of the 
phase-diagram, the two species A and B segregate into
coarsening domains of typical size $L(t)$.
The simplest model for the description of the dynamical behavior of 
these systems is the Ising model evolving with Kawasaki 
A-B exchange rule~\cite{Gunton83}.
Despite a great deal of work, mostly
numerical~\cite{Rao76,Lebowitz82,Huse86,Amar88,Roland89},
the understanding of this phase-separation kinetics is
still not completely satisfactory.
For example the expected asymptotic growth $L(t) \sim t^{1/z}$ with $z=3$
due to the evaporation-condensation mechanism (bulk
diffusion)~\cite{Lifshitz61} has not been directly
observed in simulations so far, because of extremely long transients.
This poor comprehension is also in contrast with the wide recent interest
in submonolayer ordering in epitaxial systems, for which the two-dimensional
Ising model is the simplest description~\cite{Wen96}.
In real alloys direct A-B exchange is extremely
unlikely; swaps between particles and vacancies
are instead the fundamental mechanism leading towards
equilibration~\cite{Manning68}.
Investigations of this kind of dynamics in the Ising model have been
performed~\cite{Yaldram91,Fratzl94} showing, at intermediate
temperatures, the convergence towards a late stage $z=3$ exponent
to be enhanced with respect to the Kawasaki case.
However, despite these equal asymptotic exponents,
the issue of the universality of large scale properties for the two
kinds of dynamics is far from being clarified due to a poor
understanding of the basic processes governing growth.
Several mechanisms have been proposed~\cite{Binder80,Fratzl94} for both cases,
predicting a wealth of preasymptotic exponents, but no clear
identification of them has been found in previous numerical studies.

In this Rapid Communication we investigate vacancy-mediated dynamics
in a wide range of temperatures and relative concentrations.
We identify long preasymptotic regimes during which the growth
exponent has well defined values different from the value $1/3$ expected
for the evaporation-condensation mechanism.
In particular, at low temperatures, the growth exponent is larger than
at intermediate temperatures and, for equal volume fraction
of the two species, the remarkably fast growth law $L(t) \sim t^{1/2}$
is found.
This rich phenomenology, observed numerically, is interpreted in a
unified and consistent framework by means of simple scaling arguments.
A prominent role is played by diffusion processes along the edges
of segregating domains.
This mechanism couples to the rough or faceted nature of the interfaces
at different temperatures and to the domain morphologies
at different concentrations, to give rise to a variety of growth laws.
The fact that particle-exchange is mediated by vacancy motion turns out
also to be crucial, producing a notably different
preasymptotic pattern with respect to the Kawasaki dynamics.

We consider an Ising-type model with Hamiltonian
$H=-J \sum_{<ij>} s_i s_j$ where $s_i=\pm 1,0$ 
corresponds to $A$ or $B$ particles and holes, respectively, and
the interaction is between nearest neighbors.
The only microscopic motion allowed is the exchange between a vacancy and
one of the nearest neighbor particles.
At time zero the system is instantaneously brought from an initially
high-temperature uncorrelated state to the final temperature $T<T_c$.
We perform Monte Carlo simulations of the model evolution
with the Metropolis algorithm on a square lattice of size $512 \times 512$
with one single vacancy. This hole density is
in the range of typical values for real alloys~\cite{Wagner91}.
Time is measured as the number of attempted vacancy steps.
We calculate the characteristic domain size $L(t)$ as the position of the first
zero of the correlation function averaged along the vertical and horizontal
axes.
Other measures of the domain size, as the inverse of the first moment
of the radially averaged structure factor, give similar results.
The behavior of $L(t)$ for symmetric quenches is shown in Fig.~\ref{Fig1};
here and in the following temperatures are rescaled by the critical 
temperature $T_c=2.269\ldots J/k_B$ \cite{nota}.
For very short times the vacancy
has not yet visited all parts of the lattice.
This causes the unusual behavior of $L(t)$ for $t \lesssim 10^8$.
For longer times the evolution varies remarkably depending on temperature.
At $T=0.6$, the domain size quickly starts growing as
$t^{1/3}$~\cite{Fratzl94}.
At low temperature ($T=0.2$) the effective exponent
becomes larger than $1/3$ and settles to $1/2$ during the last decade
considered.
For intermediate temperatures an even more surprising result occurs.
The value of $L(t)$ for $T=0.45$ becomes, after about $10^9$ steps,
larger than the same quantity for $T=0.6$. On much longer times, the
same occurs also for $T=0.3$. This result is at odds with the common
wisdom according to which the higher the temperature, the faster the
separation process.

For asymmetric quenches, such that the fraction of $A$ particles is $80\%$,
results are illustrated in Fig.~\ref{Fig2}.
Again for low temperature the effective growth exponent is larger
than for higher $T$ in the range of times considered.
In particular we find that at $T=0.6$, $1/z \approx 0.27$ while, at
$T=0.3$, $1/z$ is very close to 1/3.

In the following we present scaling arguments explaining the observed
behaviors in a unified framework. The basic ingredient is that,
in the time range considered, diffusion of particles along domain interfaces 
is the relevant process. This physical mechanism alone allows a consistent 
interpretation of all the phenomenology.
Our strategy to determine the
dynamical exponent $z$ is the following:
We consider first a droplet of $A$ particles
of typical size $L$ deformed from its equilibrium shape.
By denoting with $t_{eq}$
the time needed to relax the perturbation we show that
$t_{eq}\sim L^\beta$ and determine $\beta$ for the different
temperatures considered in the simulations. Then we deduce a
relationship between $\beta $ and $z$ both for symmetric and asymmetric
compositions. 

{\bf Relaxation time of a droplet deformation.}
The exponent $\beta$ depends on the roughness of the interfaces which,
in turn, is determined by temperature and the scale considered.
The roughening temperature 
for the $d=2$ Ising model is zero and therefore interfaces are always
rough on large scales. However, domain walls appear faceted at finite $T$
on length scales smaller than a typical length $L_c(T)$ which can be
estimated to be of order $\exp(2J/T)/2$~\cite{Jensen99}.
For sufficiently low temperatures, then,
$L(t)<L_c(T)$ in an appreciable time interval and interfaces are straight
on the characteristic scale $L(t)$.
This is exactly what occurs in our simulations at $T=0.2$.
For such a temperature, $L_c(T=0.2) \approx 41$ and, for all times considered
in Fig.~\ref{Fig1} $L(t) < L_c(T)$. When $T=0.6$, instead, $L_c(T) \approx 2$
and in the phase-separation process domain walls are always rough. 
We consider the case with flat interfaces first.
The basic process is the following (Fig.~\ref{Fig3}): 
A vacancy is captured on the outer part of the droplet 
interface and wanders 
along it until it switches to the inner part by moving through
a kink located at $P$. 
A passage through another kink in $Q$ and a successive 
evaporation bring the hole in the bulk of the $B$ phase again. 
The net effect is the transfer
of an $A$ particle from $Q$ to $P$. Since the vacancy moves without
energy costs along the facets, the duration of the process is 
determined by the activated events (evaporation and passage through
kinks) which take a time $\tau \sim \exp (2 J/T)$.  
Starting with a rectangular shape of size $(L-\delta) \times (L+\delta)$,
with $\delta\propto L$ ($\delta/L$ small), 
$n \simeq L\delta$ particles must be displaced from long to
short edges in order to relax the deformation. This takes a
time proportional to $n^2$, the motion being diffusive.
Then we have $t_{eq}\sim n^2 \tau \sim \exp (2J/T) L^4$, namely 
\begin{equation}
\beta=4.
\label{betalow}
\end{equation}
The equilibrium shape is then stabilized by a more favorable surface to
volume ratio. We have computed the value of $\beta$ by means of
numerical simulations of the equilibration of single droplets with
vacancy dynamics. Defining $R_x$ and $R_y$ as the gyration
radii along the $x$ and $y$ direction respectively,
$t_{eq}$ is identified with the time needed for an
initially rectangular droplet to recover the isotropic shape,
namely $R_x(t_{eq})/R_y(t_{eq})=1$.
The results at $T=0.2$ for droplets smaller than
$L_c(T=0.2) \approx 41$, shown in
Fig.~\ref{Fig4}, are consistent with Eq.~(\ref{betalow}).

For rough interfaces one can instead consider an analytic profile
and invoke the classical result by Mullins that the equilibration
time is proportional to $L^4$~\cite{Mullins63}.
However, in the present context this result must be modified to take
into account the fact that the vacancy spends part of the time
in the bulk of domains and during such intervals no dynamics occurs
on the interfaces.
Then we have $t_{eq}\simeq L^4/p$,
where $p$ is the fraction of time a vacancy stays on domain walls.
At low temperatures $p\approx 1$
(for this reason $p$ has not been taken into account when dealing with
flat interfaces)
whereas $p\simeq \exp (2 J/T ) L^{-1}$ with rough walls.
In summary we obtain  
\begin{equation}
\beta=5.
\label{betahigh}
\end{equation}
Numerical simulations at $T=0.6$
(Fig.~\ref{Fig4}) are in good agreement with Eq.~(\ref{betahigh}).

{\bf Relationship between $\beta $ and $z$.}
The vacancy-mediated phase-separation changes the relationship
between $\beta$ and $z$ with respect to Kawasaki dynamics.
The crucial observation is that only one vacancy
acts on ${\cal N}$ features of size $L(t)$.
As the typical size grows the number of objects to be relaxed
is reduced and this speeds up the separation process.
The connection between $t_{eq}$ and the growth exponent $z$ is different
for equal and unequal compositions.
In the symmetric case domain
morphology is interconnected and the kinetics proceeds
via the motion of particles along the edges away from more curved regions.
A feature of size $L$ is relaxed to equilibrium in a time
$t_{eq}\sim L^\beta$, provided the vacancy acts on it. However, since
${\cal N}\sim L^{-2}$ such features must be equilibrated by a single hole,
the time needed to relax them all is ${\cal N}t_{eq}$.
This relaxation causes an increase of the typical size $\Delta L \sim L$.
Then $dL(t)/dt \simeq \Delta L/({\cal N}t_{eq}) \sim L^{3-\beta}$ and
hence
\begin{equation}
z=\beta-2
\label{zetabetacrit}
\end{equation}
Using Eqs.~(\ref{betalow}) and (\ref{betahigh})
we obtain $z=2$ and $z=3$ for low and intermediate temperatures,
consistently with the results reported in Fig.~\ref{Fig1}.
Notice that the exponent $z=3$ observed at $T=0.6$
could be interpreted in terms of the usual evaporation-condensation 
as well.
However, this mechanism alone is not consistent with the exponent 
observed at the same temperature in the quench with unequal compositions.

In the asymmetric case the minority phase forms non-percolating
domains dispersed in a matrix of the other component.
In this case growth proceeds
via diffusion and coalescence of entire droplets. From
the knowledge of $\beta $ one can simply evaluate the diffusivity 
$D$ of a domain of size $L$ produced by particles moving along its edges.
The relaxation of a droplet to equilibrium
requires a number of molecules of order $L^2$ 
to travel diffusively a distance of order $L$;
the mean square displacement $x^2$ of the domain centre 
of mass is then of order $L$.
Therefore $D \sim x^2/t_{eq}\sim L^{-\alpha}$, with 
\begin{equation}
\alpha =\beta -1. 
\label{alfabeta}
\end{equation}
This relation has been checked numerically
by computing the droplet diffusivity for different $L$.
The results, shown in the inset of Fig.~\ref{Fig4}, 
are consistent with Eq.~(\ref{alfabeta}) for low temperature.
For $T=0.6$ the agreement is poorer, presumably because
evaporation and condensation events contribute also to the motion
of the droplet.
The exponent $\alpha$ can, in turn, be related to the coarsening growth law
by a generalization of the Binder-Stauffer argument \cite{Binder74}
to vacancy-mediated dynamics:
Two droplets separated by a distance $L$ take a time 
$t_{coa} \sim L^2/D$ to coalesce. As above, since only one vacancy is active,
the time needed for the coalescence of all droplets is ${\cal N} t_{coa}$,
${\cal N} \sim L^{-2}$ being the total number of domains.
When this happens the characteristic
domain size increases of a quantity $\Delta L \sim L$.
Then, using Eq.~(\ref{alfabeta}) one has
$dL/dt\sim \Delta L/({\cal N}t_{coa}) \sim L^{1-\alpha}$ and
\begin{equation}
z=\alpha=\beta-1.
\label{zetabetaoff}
\end{equation}
With the values of Eqs.~(\ref{betalow}) and (\ref{betahigh}) we obtain
$z=3$ and $z=4$, consistently for low $T$ with the exponents in
Fig.~{\ref{Fig2}.
For $T=0.6$, $z\approx 3.6$ consistently with the value of $\alpha$
determined numerically at the same temperature, indicating that
bulk diffusion weakly contributes.
For the sake of clarity, we summarize in Table~\ref{Tab1} the exponents
derived via scaling arguments.
It must be stressed that, with the exception of the symmetric quench
with rough interfaces, all other exponents are preasymtotic
and will crossover to 1/3 asymptotically.

In conclusion, we have shown, by means of Montecarlo simulations and
scaling arguments, that 
the phase-separation of the Ising model with vacancy-mediated dynamics
exhibits well-defined preasymptotic regimes, where coarsening is governed
by interfacial processes.
This results in the unexpected feature of faster growth at lower
temperatures.
In particular, for symmetric quenches, the effective growth exponent
attains the unusually large value 1/2.
This whole phenomenology is notably different from what occurs in the
case of Kawasaki dynamics. We expect these results to hold also with larger
hole concentrations, qualitative changes occurring only when
interfaces start to be saturated by vacancies.
Our scaling arguments can naturally be extended to the three dimensional
case, but compelling numerical evidence would require
a huge computational effort. 
The vacancy-mediated dynamics is expected to differ from the Kawasaki one
also in such a case.
Moreover, since the roughening temperature $T_R$ is finite in $d=3$,
the faceted nature of interfaces will influence even the asymptotic 
properties of the coarsening process for $T < T_R$.

We thank A. Baldassarri for a critical reading of the manuscript.
C. C. gratefully acknowledges support from the
Alexander von Humboldt Foundation.
F. C. acknowledges support by the TMR network contract ERBFMRXCT980183
and by INFM (PRA-HOP 1999).
F. C. is grateful to M. Cirillo, R. Del Sole and M. Palummo 
for hospitality in the University of Rome.

\begin{table}
\begin{center}
\begin{tabular}{||c|c|c||} \hline
Interfaces & Symmetric quench & Asymmetric quench \\ \hline
Faceted & 1/2 & 1/3 \\
Rough & 1/3 & 1/4 \\
\hline
\end{tabular}
\end{center}
\caption{Value of the growth exponent $1/z$ for the different types of
quench.}
\label{Tab1}
\end{table}

\begin{figure}
\centerline{\psfig{figure=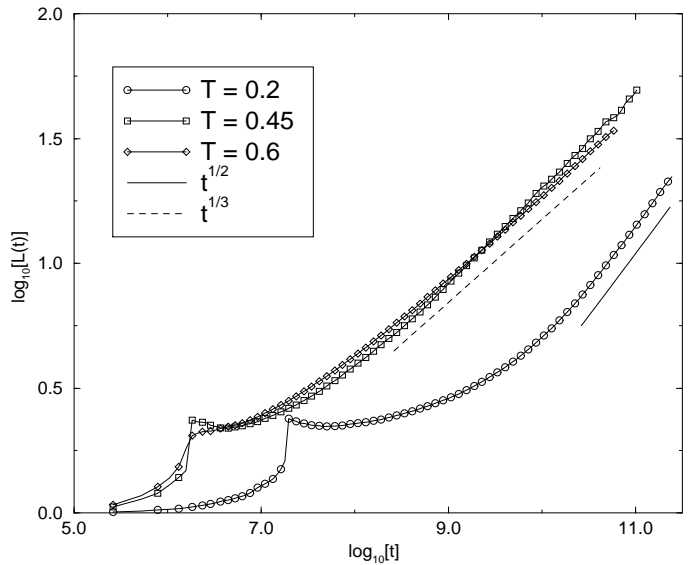,width=9cm,angle=-90}}
\caption{Double logarithmic plot of $L(t)$ vs $t$ for symmetric quenches
at various temperature. Data are averaged over 20 realizations.}
\label{Fig1}
\end{figure}

\begin{figure}
\centerline{\psfig{figure=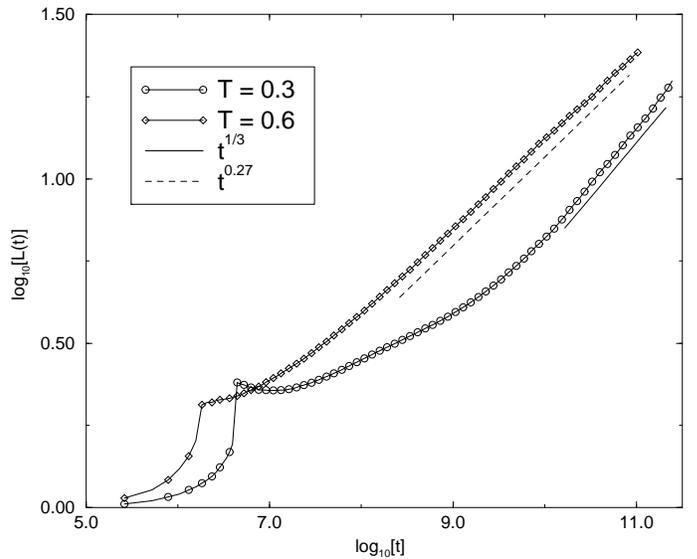,width=9cm,angle=-90}}
\caption{Double logarithmic plot of $L(t)$ vs $t$ for quenches with
unequal compositions at low and intermediate temperatures.
Data are averaged over 20 realizations.}
\label{Fig2}
\end{figure}

\begin{figure}
\centerline{\psfig{figure=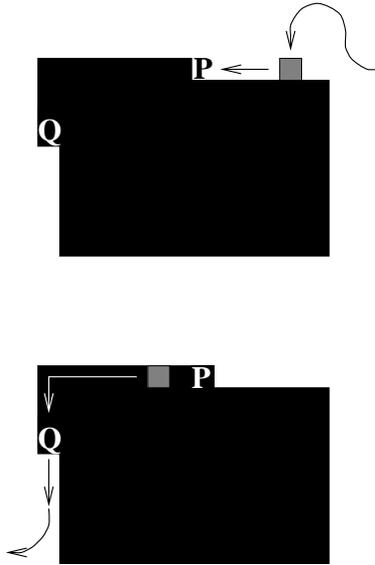,width=5cm,angle=0}}
\caption{Schematic representation of the fundamental process leading to
relaxation of a faceted droplet.}
\label{Fig3}
\end{figure}

\begin{figure}
\centerline{\psfig{figure=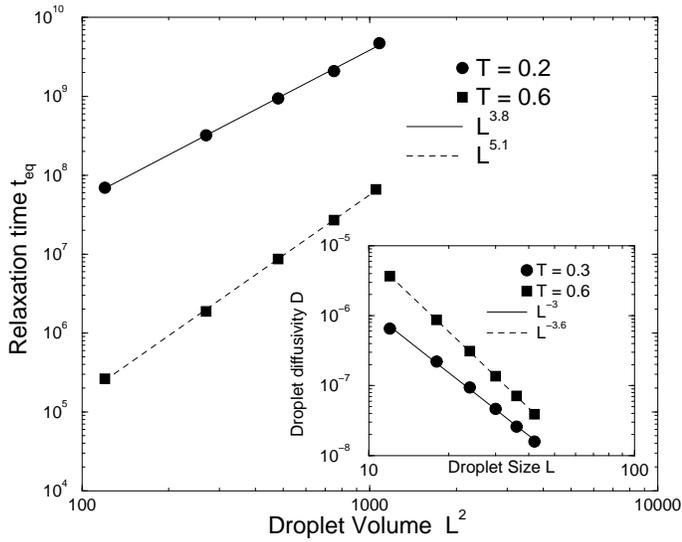,width=9cm,angle=-90}}
\caption{Main: Size dependence of the time $t_{eq}$ needed to equilibrate
an initially deformed droplet.
Best fits yield $\beta=3.8$ and $\beta=5.1$ at $T=0.2$ and $0.6$ respectively.
Inset: Size dependence of the diffusivity $D$ of a single droplet.
Best fits yield $\alpha=3$ and $\alpha=3.6$ at $T=0.3$ and $0.6$
respectively.}
\label{Fig4}
\end{figure}
 
\end{document}